\documentclass[12pt]{iopart}

\usepackage{cite}

\expandafter\let\csname equation*\endcsname=\relax 
\expandafter\let\csname endequation*\endcsname=\relax 
\usepackage{amsmath,amssymb,amsthm}

\usepackage{algorithmic}
\usepackage{graphicx}
\usepackage{array}
\usepackage{standalone}
\usepackage{etoolbox}

\newtoggle{arXiv}
\newtoggle{bibrun}

\usepackage{color, calc}
\usepackage{array, booktabs}
\newcolumntype{L}{>{$}l<{$}}
\newcolumntype{C}{>{$}c<{$}}
\newcolumntype{R}{>{$}r<{$}}

\usepackage{tikz, tikz-3dplot}
\usetikzlibrary{arrows, arrows.meta, calc, positioning}
\usetikzlibrary{decorations.pathmorphing}	
\tikzset{>=Stealth}

\usepackage{pgfplots}
\usepackage{siunitx, physics}
\usepackage{url}
\usepackage{subfigure}
\usepackage{eurosym}

\begin{document}

\title{Elliptical Halbach magnet and gradient modules for low-field portable MRI}

\author{Fernando~Galve\textsuperscript{1}, Eduardo~Pall\'as\textsuperscript{1}, Teresa~Guallart-Naval\textsuperscript{1}, Pablo~Garc\'ia-Crist\'obal\textsuperscript{1},  Pablo~Mart\'inez\textsuperscript{2}, Jos\'e~M.~Algar\'{\i}n\textsuperscript{1}, Jose~Borreguero\textsuperscript{3}, Rub\'en~Bosch\textsuperscript{1}, Francisco~Juan-Lloris\textsuperscript{2}, Jos\'e~M.~Benlloch\textsuperscript{1}, Joseba~Alonso\textsuperscript{1}}

\address{$^1$ MRILab, Institute for Molecular Imaging and Instrumentation (i3M), Spanish National Research Council (CSIC) and Universitat Polit\`ecnica de Val\`encia (UPV), 46022 Valencia, Spain}
\address{$^2$ PhysioMRI Tech S.L., 46024 Valencia, Spain}
\address{$^3$ Tesoro Imaging S.L., 46022 Valencia, Spain}

\ead{fernando.galve@i3m.upv.es}

\maketitle

\begin{abstract}

\emph{Objective.} To develop methods to design the complete magnetic system for a truly portable MRI scanner for neurological and musculoskeletal (MSK) applications, optimized for field homogeneity, field of view (FoV) and gradient performance compared to existing low-weight configurations.

\emph{Approach.} We explore optimal elliptic-bore Halbach configurations based on discrete arrays of permanent magnets. In this way, we seek to improve the field homogeneity and remove constraints to the extent of the gradient coils typical of Halbach magnets. Specifically, we have optimized a tightly-packed distribution of magnetic Nd$_2$Fe$_{14}$B cubes with differential evolution algorithms, and a second array of shimming magnets with interior point and differential evolution methods. We have also designed and constructed an elliptical set of gradient coils that extend over the whole magnet length, maximizing the distance between the lobe centers. These are optimized with a target field method minimizing a cost function that considers also heat dissipation.

\emph{Main result.} We have employed the new toolbox to build the main magnet and gradient modules for a portable MRI scanner designed for point-of-care and residential use. The elliptical Halbach bore has semi-axes of 10 \& 14\,cm and the magnet generates a field of 87\,mT homogeneous down to 5,700\,ppm (parts per million) in a 20\,cm diameter FoV, it weighs 216\,kg and has a width of 65\,cm and a height of 72\,cm. Gradient efficiencies go up to around 0.8\,mT/m/A, for a maximum of 12\,mT/m with in 0.5\,ms with 15\,A \& 15\,V amplifier. The distance between lobes is 28\,cm, significantly increased with respect to other Halbach-based scanners. Heat dissipation is around 25\,W at maximum power, and gradient deviations from linearity are below 20\% in a 20\,cm sphere.

\emph{Significance.} Elliptic-bore Halbach magnets enhance the ergonomicity and field distribution of low-cost portable MRI scanners, while allowing for full-length gradient support to increase the FoV. This geometry can be potentially adapted for a prospective low-cost whole-body technology.
\end{abstract}

\section{Introduction}
Low-field Magnetic Resonance Imaging (MRI) systems can provide clinical diagnostic potential in situations inaccessible to standard scanners \cite{Deoni2022,GuallartNaval2022,Algarin2023}, as field strengths below 100\,mT have been shown to suffice to diagnose relevant medical conditions \cite{Sheth2021,Sheth2022,Liu2021}. Consequently, research and clinical activity in this area are rapidly escalating and promise a dramatic increase in MRI accessibility globally \cite{Sarracanie2020,Wald2020,Webb2023}. 

In this regime, two usual magnet arrangements to polarize the hydrogen nuclei in the patients' tissues are yoked magnets and Halbach arrays. The former can produce highly homogeneous fields but are relatively heavy due to the massive iron parts that guide the magnetic field lines, see e.g. He {\it et al.} (51\,mT, 120\,ppm, 350\,kg, \cite{chonqing2020}), Sheth {\it et al.} (64\,mT, 630\,kg, \cite{Sheth2021}), or Liu {\it et al.} (55\,mT, 250\,ppm, 750\,kg, \cite{Liu2021}). On the other hand, sparse Halbach mandhalas \cite{Soltner2010}, i.e. collections of small magnetic pieces in a Halbach configuration, have proved a great tool to produce light and inexpensive MRI magnets, but they are usually less homogeneous, see e.g. O'Reilly {\it et al.} (50\,mT, 2,500\,ppm, 75\,kg, \cite{OReilly2020}), or Guallart-Naval {\it et al.} (72\,mT, 3,100\,ppm, 250\,kg, \cite{GuallartNaval2022}). Note that the definition of homogeneity varies across the literature and precludes a direct comparison between different systems, but the above figures are consistent with the fact that magnetic yokes can be designed to tailor the field within a region of interest.

\begin{figure}
\includegraphics[width=\textwidth]{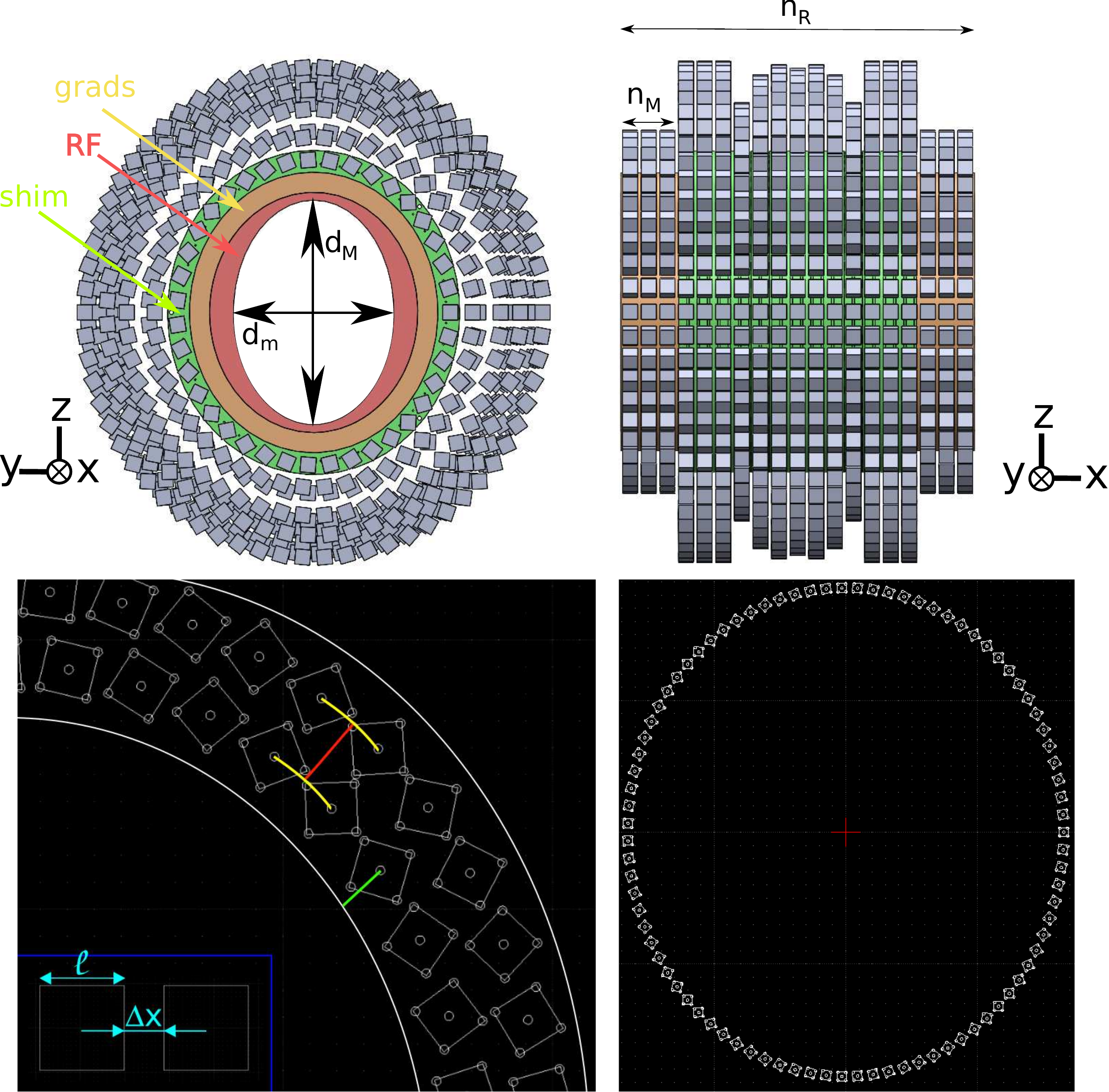}
\caption{Sketch of generic magnet. (Top-left) The bore has minor and major diameters ($d_\text{m}$, $d_\text{M}$), and radially outwards we find the RF holder, the gradients, and the shimming unit. (Top-right) The magnet has a number of outer (`mouth') rings $n_\text{M}$, within a total of $n_\text{R}$ rings, where the gradients extend full length $L$ (i.e. they are surrounded by $n_\text{R}$ rings), while the shimming unit only extends the central part (i.e. $n_\text{R}-n_\text{M}$ rings). (Bottom-left) An example ring with two layers ($n_\text{L}=2$, and cubes of side $l$. (Bottom-right) Shimming unit rings have equal dimensions and number of slots, corresponding to the green support in upper figures.}
\label{fig:definitions}
\end{figure}

Magnets based on Halbach arrays are usually conceived with a cylindrical bore, which naturally produces an elongated field pattern \cite{Soltner2010}. This can be compensated with an elliptical design, which is also advantageous in terms of accommodating the patients' limbs and heads. Unfortunately, the literature on Halbach magnets with elliptical bores is scarce so far, and mostly focused on charged particle beam shaping \cite{gluckstern1983,lund1996,kustler2010}. An exception is an expired patent disclosing an elliptical-bore Halbach magnet for full-body MRI \cite{gluckstern1996}. While they follow the idea of stacking rings (as the design concept in Fig.\,\ref{fig:definitions}), they do not use individual magnetic pieces {\it à la mandhala}, but compact blocks made of equally-oriented bricks with their magnetic field pointing along the major ellipse axis, thereby removing degrees of freedom required for homogenizing the field. Perhaps more relevant is a recent study exploring the dependence of the Halbach rotation rule for elliptical cross-sections \cite{permanentWebb2023}. Interestingly, they provide a permanent magnet hypothesis to guide angular arrangements for arbitrary cross-sections. However, this is restricted to Halbach arrangements with a single layer of permanent magnets ($n_\text{L}=1$), which limits the ultimate field strength achievable (their simulated optimal design results in 28\,mT).

Genetic algorithms are the usual method of choice for Halbach magnet design \cite{Cooley2018,OReilly2019}, as they are well suited for multi-dimensional configuration spaces, nonlinear cost functions and integer variables (such as the number of magnetic elements in a given ring). Nevertheless, alternative methods (e.g. interior point, \cite{interior2000}) or evolutionary algorithms (e.g. differential evolution, \cite{DiffEvol1997}) are also appropriate for this task and have been largely ignored or overlooked by the low-field MRI community. Interior point methods are designed for nonlinear convex continuous variable problems, but field homogeneity can be cast into a squared deviation from ideality. Both methods are designed for continuous variables, but projection onto integers render them valid for our problem and, most importantly, they are fast. In some cases, differential evolution is known to be more efficient than genetic algorithms \cite{DEcomparative2004}, and it is becoming a research option for weight \cite{DEneural2020} and topology \cite{DEneuralTopology2022} optimization for machine learning. This speed-up in calculations makes optimization in a large number of dimensions feasible, meaning new degrees of freedom (e.g. multiple magnet sizes, variable number of layers, etc.) can be explored within reasonable computational times.

Halbach magnets in portable MRI systems tend to be short for weight considerations and to ease access to their center (see Fig.\,\ref{fig:definitions}). However, the ratio between the axial and radial dimensions defines to a large extent the homogeneity of a Halbach magnet; when this ratio is small, the magnet rings at the extremes (`mouths') must be as radially tight as possible. This often constrains the axial space available for gradients to the larger magnet rings, limiting the available imaging region due to field nonlinearities near the centers of the gradient coil lobes. One solution is to install the gradient system externally to the magnet, but this severely compromises their performance, especially in terms of achievable gradient strength, thereby affecting image resolution and precluding useful pulse sequences such as for diffusion-weighted imaging. Thus, gradient coils are usually placed inside the magnet. We seek a system just large enough to fit the head (accommodating the shoulders would make the system larger and hardly portable), so the distance between the shoulders and the magnet center needs to be short ($\approx22$\,cm, see Methods). In this case, full-length gradients (i.e. gradients that use all the longitudinal space available between the magnet openings) are desirable. Hence, the challenge is to produce a large-bore, short and homogeneous elliptical magnet that provides full-length support for the gradient assembly.

\begin{figure}
	\begin{center}\includegraphics[width=11cm]{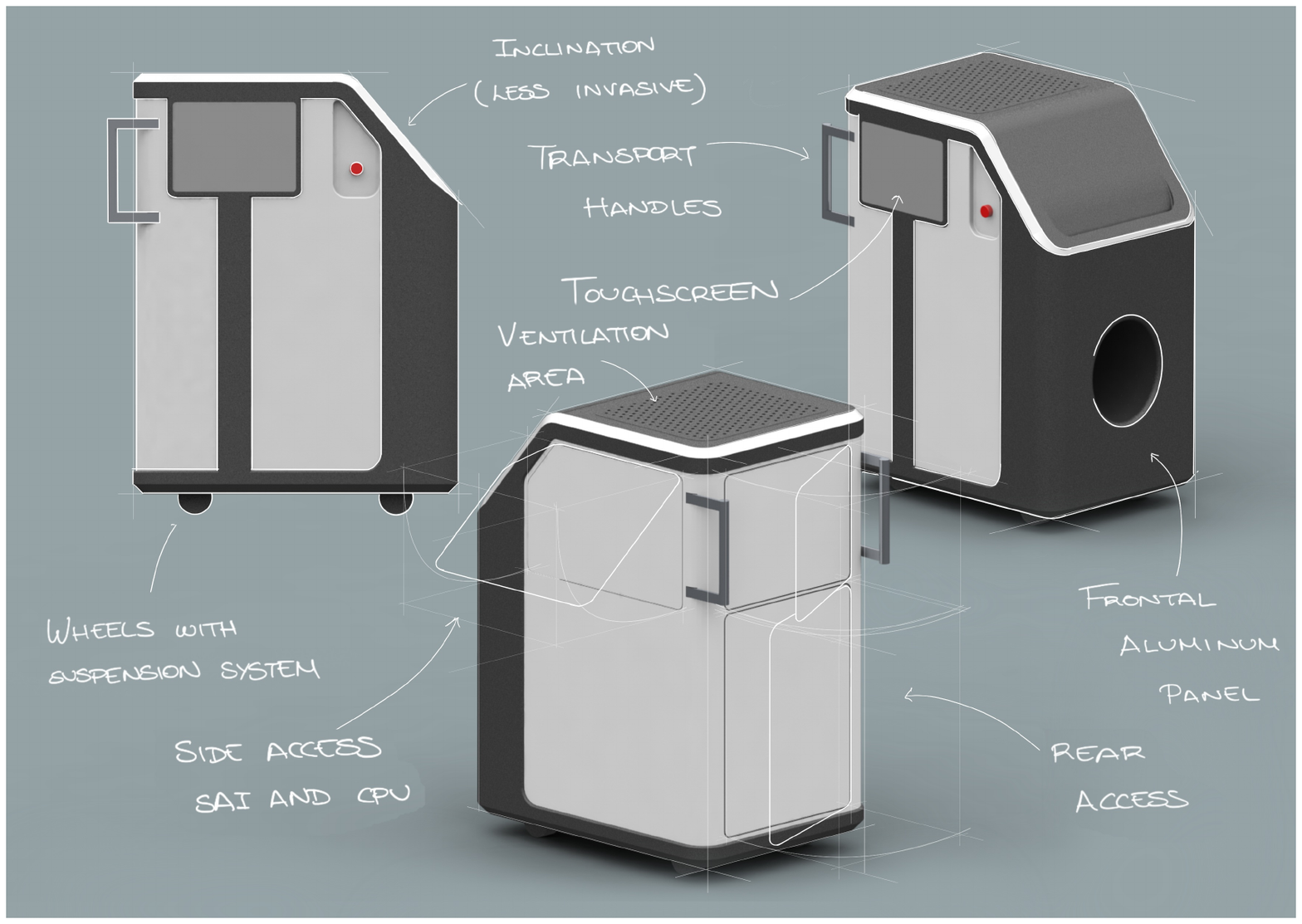}\end{center}
	\caption{Sketch of complete scanner housing.}
	\label{fig:Concept}
\end{figure}

In this paper we present a toolbox to design magnetic systems for low-field MRI scanners, including both the main magnet and the gradient coils. These tools include interior point algorithms and differential evolution algorithms for discrete Halbach magnets and shimming modules (both of which can handle elliptical cross sections), as well as basis functions for current densities on elliptic surfaces for gradient coil optimization \cite{turner86,liGrad2008}. We exemplify the use of these tools with the design, construction and characterization of the complete magnetic system for an extremity and brain MRI scanner (Fig.\,\ref{fig:Concept}) featuring high portability, small footprint (small enough to fit through standard European door clearance) and low cost ($<100$\,k€). Our prototype includes a Halbach magnet with an elliptical bore with outer diameters 65 \& 72\,cm and a length of 45\,cm, enabling neurological and musculoskeletal (MSK) applications while retaining full-length support for gradient coils. Our gradients have full-length support, providing a 28\,cm lobe-to-lobe distance and thus a potentially larger FoV, and they are made of tracks, which reduces heat dissipation by a factor $\times3-\times4$ compared to wire-based designs.

\section{Theoretical framework}
\label{sec:theory}
In this section we discuss the decision process we consider for the design and assembly of a low-field Halbach scanner. We aim at delineating the design process from the abstract scanner constraints and objectives to the point where an optimization algorithm gives the best solution.

\subsection{Decision tree for scanner constraint determination}
When building a scanner Halbach from scratch, designers run over an implicit abstract decision tree which determines the starting point for the process. Some of its key elements can include:
\begin{itemize}
 \item Is it portable?
 \begin{itemize}
	\item Yes: Depending on whether the scanner is to be motorized or not, there is a higher or lower, but indeed stringent, constraint on {\it weight}.
	\item No: No strict weight limit, or limited by architectural and/or cost constraints.
 \end{itemize}
 \item Point-of-care, residential or indoor use?
 \begin{itemize}
	\item Yes: The maximum dimensions (notably the footprint) are determined by door clearances or the use of elevators.
	\item No: No strict limit on outer dimensions, or limited by architectural and/or cost constraints.
  \end{itemize}
 \item Imaging application?
 \begin{itemize}
	\item Full body: longer FoV, and higher eccentricity.
	\item Extremity: shorter FoV, lower eccentricity for arms and higher for legs/feet.
	\item Head: Depending on limitations on total weight and dimensions, one may consider accommodating the shoulders in the scanner (longer magnet, lower field, but better homogeneity), and thus the eccentricities change accordingly. If shoulders are to remain outside the scanner, an asymmetric configuration can be chosen. i.e. with the FoV center displaced axially from the magnet center.
  \end{itemize}
  \item Shimming
 \begin{itemize}
	\item Outside/inside magnet?: this decision affects the geometric constraints for magnet optimization.
	\item Active/passive?: active shimming with an extra set of coils (i.e. not the gradients) allows for dynamic stabilization with a feedback loop but can complicate design, assembly and operation, more so if the shimming module is inside the main magnet.
  \end{itemize}
 \item Gradients
 \begin{itemize}
	\item Outside/inside magnet?: Outer gradients provide more freedom for design and heat management, but are larger and thus less efficient. Inner gradients are more efficient but harder to assemble and cool down.
	\item Full length?: When gradients are within the magnet, full length (i.e. at least as long as the magnet) imposes geometric restrictions on the magnet, but can provide larger inter-lobe separation, increasing the available FoV. Furthermore, cable connections can be easier to manipulate for full-length gradients. On the other hand, shorter gradients can feature lower inductances (i.e. faster rise times) and operate at reduced power, in exchange for a smaller FoV.
 \end{itemize}
 \item Performance
 \begin{itemize}
  \item Field strength: Even after fixing the outer dimensions and total weight, there is still some freedom to choose the field strength. Intense fields typically require dense arrangements of magnetic material, which leaves less room to optimize for homogeneity.
  \item Gradients: Geometric constraints impose strong limitations on inductance, heat dissipation and rise times, which can be insufficient for certain sequences such as diffusion weighted\cite{QIMS5938} or functional imaging\cite{BOLLMANN2021}.
  \item Shimming: A bad homogeneity can lead to a short $T_2^*$ or unmanageable off-resonance artifacts. To improve the homogeneity, the shimming module can include extra layers of permanent magnets (passive) or coils (active).
  \end{itemize}
\end{itemize}

Once the above items are analyzed and decided upon, there is a process of optimization where a balance is sought between the target specifications. For portable MRI, the system dimensions and total weight are hard constraints, but field strength and homogeneity are not, so one can pursue a dense configuration of magnetic material at the expense of difficulty in assembly, or choose larger/smaller magnetic elements with more or less radial layers. The same applies for the shimming unit, whose spatial support restricts the freedom of field homogeneity optimization. Finally the gradients can be chosen as wires, which typically dissipates more heat and requires a mechanical support with patterned grooves (even if flexible Litz wires are sometimes considered \cite{litzWireGrad2021}), or a metal plate laser-cut or water-jetted to form the coil/lobe. The latter is significantly more demanding in terms of design optimization, manufacturing and assembly, but otherwise performs better (see previous section). Another consideration when opting for gradient coils made out of cut metal plates is that they may leave less room for field shimming.

\subsection{Magnetic optimization approach}
\label{theo:opt}
For the main field and shimming units we approximate magnetic elements to behave as ideal dipoles and we ignore coercitivity effects, simplifying the problem by assuming linear superpositions of dipole contributions. We have carried out multiple simulations including realistic shapes and magnetic properties, and their influence is systematically negligible compared to the deviations we observe in real assemblies due to machining, assembly and magnetization tolerances. The magnitude of the equivalent magnetic moment of a magnetic element is given by $m=B_\text{rem} V/\mu_0$, with $B_\text{rem}$ the remanence, $V$ the volume and $\mu_0$ the permeability of vacuum. The dipolar approximation is quite good at distances greater than twice the magnetic piece's size \cite{dipole2013}. Coercitivity effects complicate significantly the optimization problem, and therefore how fast and efficiently the configuration space can be searched. Thus, we neglect them during the optimization stage, but their effect is later ascertained by finite-elements simulation in COMSOL Multiphysics (Stockholm, Sweden).

\subsection{Basis functions for elliptic gradient supports}
\label{Basis}

For an elliptical Halbach magnet, the gradient arrangement that most naturally adapts to the available space is to place the current-carrying elements on curved surfaces with an elliptical cross-section. We do not consider alternative geometries which may require methods based on e.g. stream functions defined as finite elements on an arbitrary support surface. Instead, the below formalism generalizes the stream-function approach typically employed to optimize gradient geometries on circular cross-sections \cite{OReilly2020} to elliptical cylindrical surfaces. It therefore serves as basis for the optimization of elliptical gradients, and we have not found an equivalent elsewhere in the literature.
%

We use the coordinate convention $\{x,y,z\}$ shown in Fig.\,\ref{fig:definitions}, with $y=a\sinh\rho\sin\phi$, $z=a\cosh\rho\cos\phi$, and $a$ is the distance between ellipse {\it{foci}} and center. We define the major and minor semi-axes of the gradient support, $L_z=a\cosh\rho$ and $L_y=a\sinh\rho$ respectively (note these two quantities are slightly different for each gradient), and its longitudinal extent  $L_x$.

In said elliptical coordinates, the current densities $J_\rho=0$ and $(J_x, J_\phi)$ are related by the continuity equation
\begin{equation}\label{eq:continuity}
	\vec{\nabla}\cdot \vec{J}=0,
\end{equation}
so we can focus on $J_\phi$. The expressions for the basis functions for each gradient coil (which we label as GX, GY, GZ) are:
\begin{eqnarray}
	\text{GX:\,} & J_\phi=\sum_{n,m} P_{n,m}\sin(n\phi)\cos\left(\frac{m\pi}{L_x}x\right) & \text{n odd, m even},\\
	\text{GY:\,} & J_\phi=\sum_{n,m} P_{n,m}\cos(n\phi)\sin\left(\frac{m\pi}{L_x}x\right) & \text{n even, m odd},\\
	\text{GZ:\,} & J_\phi=\sum_{n,m} P_{n,m}\sin(n\phi)\sin\left(\frac{m\pi}{L_x}x\right) & \text{n even, m odd}.
\end{eqnarray}

From here on we present the calculations for GY, as for the rest we may follow a similar procedure. To calculate the field produced by the surface current densities we require knowledge of $J_x$ only, since $J_\rho=0$. From Eq.\,(\ref{eq:continuity}) we get
\begin{equation}
	J_x=-\frac{1}{h}\partial_\phi{J_\phi} - J_\phi\frac{a^2\sin\phi\cos\phi}{h^3},
\end{equation}
with the scale factor $h=h_\rho=h_\phi=a\sqrt{\sin^2\phi+\sinh^2\rho}$ (and $h_x=1$), i.e.
\begin{equation}
	J_x=\sum_{n,m} P_{n,m}\left[\frac{-nL_x}{hm\pi}\sin(n\phi)\cos\left(\frac{m\pi}{L_x}x\right)+\frac{L_x}{m\pi}\cos(n\phi)\cos\left(\frac{m\pi}{L_x}x\right)\frac{   a^2\sin\phi\cos\phi}{h^3} \right].
\end{equation}

The main component of the magnetic field is $B_y$ with the convention in Fig.\,\ref{fig:definitions}, and this is given by Biot-Savart's law through the $J_x$ and $J_z$ components. To find $J_z$ we can decompose the vector $\vec{J}_\phi= J_\phi \hat{\phi}$, by expressing $\hat{\phi}=\partial_\phi{\vec{r}}/|\partial_\phi{\vec{r}}|$ and $\vec{r}=x\,\hat{x}+a\sinh{\rho}\sin{\phi}\,\hat{y}+a\cosh{\rho}\cos{\phi}\,\hat{z}$, so that $\hat{\phi}=(a\sinh{\rho}\cos{\phi}\,\hat{y}-a\cosh{\rho}\sin{\phi}\,\hat{z})/h$. Then:
\begin{equation}
	J_z=\frac{-L_z\sin\phi}{h}\sum_{n,m} P_{n,m}  \cos(n\phi)\sin\left(\frac{m\pi}{L_x}x\right).
\end{equation}
Finding $B_y$ requires knowing $(\vec{J}\times\vec{r})_y$, where $\vec{r}$ is the distance between the current element and the center of the FoV. We parameterize current positions as $(x,L_y\sin\phi,L_z\cos\phi)$. Currents can be integrated over the complete elliptical surface $$\phi\in[0,2\pi], \,\,x\in[-L_x/2,L_x/2],$$ yielding $(B_y)_i\equiv B_y(\vec{r}_i) =\sum_{n,m}P_{n,m} M_{n,m}(\vec{r}_i)$ at the chosen FoV points $\vec{r}_i$, or simply $B_y=M^T P$, as usual (when programming, the indices $n,\, m$ are vectorized to a single index). It is also customary to restrict the maximum allowed $n,\,m$, since higher orders are known to produce higher current densities which yield higher heat dissipation, even if the resulting gradient fields can become more linear.

The heat dissipated by the current density distribution over the surface $S$ is
\begin{equation}
	P_\text{diss}=\frac{\delta}{t}\int_S (J_x^2+J_\phi^2)h\, d\phi dx,
\end{equation}
with $\delta$ the resistivity of the material and $t$ the track thickness. A similar procedure for this integral results in the matrix equation $P_\text{diss}=P^T\, G\, P$.

We here follow a standard target method where current densities try to produce a linear field $B_\text{target}$ in the FoV \cite{turner86}. A cost term is included to constrain power dissipation within acceptable values \cite{liGrad2008}. The cost function to be optimized is then
\begin{equation}
	{\cal{F}}=\sum_i [(B_y)_i - B_{\text{target},i}]^2 + \lambda P_\text{diss}=||M^T P- B_\text{target}||_2^2 +\lambda P^T\, G\, P
\end{equation}
which is extremal when $\partial{\cal F}/\partial P=0$, i.e. when
\begin{equation}
	P=(2MM^T+\lambda G)^{-1}(2MB_\text{target}).
\end{equation}

After optimization, tracks can be defined as usual from the stream function, which we call $\vec{\Psi}$\cite{streamFunc}. Since current densities are $\vec{J}=\vec{\nabla}\times\vec{\Psi}$, and $\vec{\Psi}$ is normal to the surface (i.e. it points along $\hat{\rho}$), we have that $J_\phi=\partial_x\Psi$ and
\begin{equation}
	\Psi=-\sum_{n,m} P_{n,m}\cos(n\phi)\cos\left(\frac{m\pi}{L_x}x\right)\frac{L_x}{m\pi},
\end{equation}
which can be obtained from the optimized coefficients $P_{n,m}$.

\section{Methods}
In the following, we describe our use of the framework introduced in the previous section to design a portable MRI scanner based on an elliptical Halbach magnet.

\subsection{Determination of scanner constraints}

\subsubsection{Main system constraints.}
Our aim is to design and build a non-motorized portable scanner for neurological and MSK applications, to be used in hospitals, primary care centers, small and specialty clinics and point-of-care and residential use. As such, it needs to comply with restrictive weight constraints and fit through standard door clearances. For these uses, the scanner must be loaded onto elevators, emphasizing the need for low weight and limiting also the total length and height. Thus, we have decided for configurations that leave out the patients' shoulders. System simplicity, industrializability and modularity are also sought.

Since optimizing the homogeneity of the main field of a Halbach unit invariably results in ``mouths'' that ultimately limit the magnet bore dimensions (Fig.\,\ref{fig:definitions}), we choose to use this space inside the central rings to place the shimming unit.

For neurological applications with a magnet which does not cover the shoulders, the distance between the brain and the bore opening is small. Consequently, we decided for gradients that have full-length support; otherwise the inter-lobe separation would limit the linear gradient region, severely compromising the size of the FoV along the axial direction ($x$). 

In terms of cost and performance, we aim at similar criteria as those used in a previous portable system \cite{GuallartNaval2022}.

\subsubsection{Dimension constraints imposed by clinical applications.} \label{sec:CriteriaApp}
Our target specifications for the scanner are:
\begin{enumerate}
	\item large enough for feet (non-flexed) of European size 46/47 and human heads  up to the 99$^\text{th}$ percentile;
	\item gradient coils spanning the full magnet length; and
	\item system small enough to fit through standard door clearance (70\,cm maximum breadth).
\end{enumerate}
Feet with size 46/47 span over 30\,cm in the longest dimension. The system bore opening need not be that large, though, as the ankle can be inserted at an angle even without flexing the possibly injured joint. For knee imaging, the thigh needs to fit in the scanner. Since a potential application niche for our portable scanner is in sports clubs and events and the thigh perimeter of some sportspeople can be $>55$\,cm, we constrained the $y$ opening to be $\geq20$\,cm. On the other hand, human male heads in the 95$^\text{th}$ percentile have an ear-to-ear distance below 21\,cm and pronasale to back of the head span of at least 21\,cm \cite{humano}. Thus, we constrained the $z$ opening to be $\geq24$\,cm, preferably closer to 30\,cm to have some leeway.

\subsection{Main magnet}

\subsubsection{Halbach design criteria.}
Since we seek simplicity, industrializability and modularity, we have chosen a design based on magnetic rings stacked longitudinally (Fig.\,\ref{fig:definitions}). Each ring is machined out of polyethylene plates and includes sockets to accommodate N48 neodymium boron iron (NdBFe) cubic magnetic elements of side $l$. Cubes are oriented with one face parallel to the ring surface and each ring has a number of layers distributed radially outwards. This Halbach dipolar configuration generates a field mainly along $y$, perpendicular to the longidutinal direction.

To attain a relatively high field, we aim at a tight configuration of cubes, which have been arranged as follows (see Fig.\,\ref{fig:definitions}): angularly, cube centers are separated by a distance of $\sqrt{2}l+0.6$\,mm; radially, by $\sqrt{2}l+1$\,mm, where the innermost layer has cube centers $\sqrt{2}l/2+0.6$\,mm away from the inner border of the ring; longitudinally, cubes are $(l+\Delta x)$ apart (with $\Delta x=4.4$\,mm, see Results for a discussion on the effect of modifying this distance). These constraints follow from geometrical considerations and mechanical tolerances during the machining. Note that the sockets for the magnetic elements are blind (rather than through-hole) and that all rings are sealed with aluminum covers to withstand the sizable interactions that can arise between the closest elements in neighboring rings. These covers are only 2.5\,mm thick so as not to add too significant weight to the main magnet. Note also that all rings have an extra set of aligned holes to insert long threaded rods that traverse the whole stack for assembly and mechanical stability.

\subsubsection{Symmetries and other considerations.}
\label{sec:MetSymm}
Reckoning that, while constrained, the design problem is multi-dimensional, we have considered a variety of possibilities. 
The first decision to be made is whether the ring configuration is reflection symmetric along the longitudinal direction. For example, one could aim for a FoV displaced axially with respect to the magnet center \cite{webbIsmrm,Cooley2018}. These arrangements are harder to homogenize, since the number of variables is doubled, and more sensitive to experimental imperfections, especially for short magnets. After several trials, we decided to restrict our design to be axially symmetric (see Sec.\,\ref{sec:ResSymm}).

In the realm of symmetric configurations, we explored the option to build the largest rings possible to hold as many magnetic elements as fit in a Halbach configuration, and then remove elements to optimize for field homogeneity, inspired by ideas in Ref.\,\cite{Cooley2018}. Unfortunately, this performed invariably worse than packing as many magnetic elements as possible (i.e. no empty sockets) and letting an algorithm find where to place them.

Restricting to symmetric, tightly packed configurations, one could use different cube sizes, number of layers or ellipticity for each ring. We have studied arrangements with alternating cube sizes for consecutive rings, finding no advantage. We also attempted to increase the fields strength and homogeneity with an extra layer in the mouth rings, but to no avail. After numerous simulations, we decided to use cubes of equal size and rings with the same number of layers.

For a given magnet length, the size of the cubic magnetic elements defines the number of rings in the stack. However, there is some freedom to optimize the homogeneity by increasing the inter-ring or inter-layer separation, albeit at the expense of field amplitude. We also explored these possibilities and opted for the tightest possible axial packing.

All these degrees of freedom have been analyzed independently in this work, but could be combined as decision variables in the optimization problem. Note also that we set the number of rings in each mouth to either two or three, with the tightest radial distribution, i.e. they were not considered as optimization variables.

\subsubsection{Magnet dimensions.}
\label{sec:MetDim}
The magnet bore dimensions must suffice to house other scanner components (shimming unit, gradient module, RF coils) and still provide ample room for the patient (see target specifications in Sec.\,\ref{sec:CriteriaApp}).

The magnet length is determined by the distance from the shoulders to the center of the brain ($\approx5$\,cm above the eyes). This is $\approx22$\,cm for males up to percentile 95\cite{humano}. This distance also suffices for children above 8 years of age (percentile 95, too) to reach the center of the FoV knee imaging\cite{humano}. Our magnet is axially symmetric, so the total length was set to 44\,cm. Note that this largely determines the achievable magnet strength and homogeneity. While a longer magnet improves the overall field quality, it forces imaging in regions of higher inhomogeneity (see Results section for a discussion on this trade-off). Note, too, that the prototypes in Refs.\,\cite{OReilly2020} and \cite{GuallartNaval2022} are both around 51\,cm long and therefore easier to homogenize.

The second requirement in Sec.\,\ref{sec:CriteriaApp} is to allow for full-length gradient support. In line with previous designs, the RF coil is the innermost element, followed by the gradient module, the shimming unit and the magnet rings, and we define a spherical FoV of diameter 20\,cm  \cite{OReilly2020,GuallartNaval2022}. Both the RF coil and shimming unit can be shorter than the magnet, but the sought full-length gradient support means that the magnetic rings in the mouths of the magnet are constrained by the outer diameter of gradients, which are in turn limited by the outer diameter of the RF coil. Furthermore, gap capacitors to suppress wavelength effects in RF coils need to withstand high voltages and can be sizable ($\sim2$\,cm), so they must be considered in this context. They can be placed along the any radial direction, and thus influence the ellipticity of the magnet rings (see Results). Considering an RF coil holder of 1\,cm radial thickness, and another 2.5\,cm for the full gradient assembly, the rings at the mouth are constrained by an extra 3.5\,cm increase in diameter, 5.5\,cm in the direction with capacitors (see Table\,\ref{table:radialWidths} below). These rings greatly influence field strength and homogeneity, so this is entered as a hard constraint to the optimization problem.

\begin{table}[]
\begin{center}
\begin{tabular}{|l|l|l|l|l|}
\hline
             & RF holder & Gap capacitors & Gradient set & Total       \\ \hline
radial width & 1\,cm      & 2\,cm / 0\,cm        & 2.5\,cm       & 5.5\,cm / 3.5\,cm \\ \hline
\end{tabular}
\end{center}
\caption{Radial widths along the minor and major axes for the different elements that lie between the patient and the magnet rings.}
\label{table:radialWidths}
\end{table}

Finally, the third condition limits to 70\,cm the horizontal dimension of the largest rings, and therefore affects the size of magnetic elements, as well as the total number of layers. On the other hand, a small number of layers favors low-weight and the contribution of magnets far from the center of the FoV is small. So constraining to 70\,cm outer dimension is a reasonable compromise in terms of field strength and homogeneity, and system weight. Naturally, we choose to place the magnet such that it fulfills the 70\,cm condition in the shorter dimension $y$, i.e. the major axis is along $z$.

\subsubsection{Magnet optimization and computation.}
\label{sec:MetOpt}
For optimization of the field produced by the elliptical Halbach magnet we consider magnetic cubes in the dipole approximation, with no coercitivity effects and with an average remanence $B_\text{rem}=1.4$\,T (N48 grade). The optimization variables are the number of cubes $N_i$ (we use upper case to highlight them) in the innermost layer of a given ring $i$, with each additional outer layer having 7 more cubes than the former, which satisfies the design principles and tolerances for the ranges we are considering. The lowest possible value of $N_i$ is constrained by the inner bore, i.e. it depends on whether the ring hosts only the gradients ($n_\text{G}$) or includes also the shimming units ($n_\text{S}$), and the maximum possible value is constrained by the outer minor diameter of 70\,cm ($n_\text{MAX}$). The minor semi-axis of each layer is proportional to $N_i$, with the factor given by $f=(\sqrt{2}l+0.6$\,mm$)/2\pi$ (Fig.\,\ref{fig:definitions}). The cube positions are given by an ellipse with minor semi-axis $f\cdot N_i$ and major semi-axes $f\cdot N_i +\Delta R$, where $\Delta R$ depends on the major bore semi-axis and on whether gap capacitors are placed along the minor or major directions (our final configuration will have them along the minor direction because this improves homogeneity). In general, we considered axially symmetric configurations with a fixed number of mouth rings $n_\text{M}$ (two or three, depending on the cube size), i.e. $[ n_\text{G}$, $n_\text{G}$, $n_\text{G}$, $N_1$, $N_2$, ..., $N_2$, $N_1$, $n_\text{G}$, $n_\text{G}$, $n_\text{G}]$ if we have three fixed mouth rings, where variables are optimized in the range $N_i\in [n_\text{S},n_\text{MAX}]$.

The field ($B_y$) is the contribution of each cube considered as a dipole in the 20\,cm diameter FoV in the center, and the optimization cost function is the inhomogeneity $(\text{max}(B_y)-\text{min}(B_y))/\text{mean}(B_y)$. One can add a penalty to the cost function to bias the result for higher field strengths. This can be realized by a monotonically or piecewise-decreasing penalty towards the targeted field strength. We found this to work reasonably to determine field strength based on the amount of layers and the size of cubes, although a simpler method is to constrain maximum radii of rings to smaller values.

Our first trials used the DEAP Python library for genetic algorithm optimization (modified from the open repository in Ref.\,\cite{GenAlgRepo}), with a population size of 25,000, 250 iterations (0.55 crossover and 0.45 mutation, as in Ref.\,\cite{OReilly2020}). This required $\sim$4 minutes when single-threaded in an AMD Ryzen 9 7950X processor with no GPU acceleration (which could improve performance by considering more points in the FoV when calculating $B_y$). Secondly, we programmed in Julia \cite{Julia}, which offers a simple python-like syntax and near-C performance. We made use of the BlackBoxOptim.jl library (BBO, \cite{Feldt2018}), which contains multiple evolutionary algorithms. Results converged in around 30,000 iterations, which takes 5.3 seconds including library loading (5\,s otherwise) and were comparable to those with Python and DEAP. Artificially forcing the code to take as many FoV points as in Ref.\,\cite{GenAlgRepo}, it took 36 seconds, meaning that Julia is at least a factor 7 faster. Furthermore, in Julia one can easily give different optimization ranges for each variable and provide a seed configuration, making it very flexible. In addition it is trivial to incorporate multi-threading or GPU-computing into the code. We present some convergence analysis for both libraries in the Results section. There we analyze the different algorithms present in BBO for different population sizes, and conclude that the method and parameters chosen by default perform best. 

Additionally we tested the EvoLP.jl library \cite{Sanchez-DiazEvoLP2023a}, and the interior point methods contained in Ipopt.jl \cite{Wachter2006} (within the JuMP.jl framework \cite{JuMP}). However, the latter did not yield satisfactory solutions and the former was too slow.

\subsection{Shimming}
\label{sec:MetShim}

The shimming unit contains magnetic elements in the central region of the main magnet, ideally arranged so as to homogenize the field generated by the main magnet in the FoV. This unit is a also built as a stack of 23 rings. Given the limited space available, it incorporates a single layer of magnetic elements and 87 sockets per ring, as in Ref.\,\cite{GuallartNaval2022}. The optimization degree of freedom for every socket is whether to fill it with a magnet in the Halbach direction ($+1$), in the opposite direction ($-1$), or to leave it empty (0).

After assembly of the main magnet, $B_0$ was measured experimentally by scanning a THM1176 probe (Metrolab, Geneva, Switzerland) mounted on a 3-axis motorized positioning system. The field at every point $i$ was measured with 1,000 averages with a bandwidth of ~1\,kHz, 2\,s after the motors stopped to provide enough time for mechanical oscillations to fade. We scanned two Cartesian grids with 1\,cm resolution in all three dimensions, displaced 5\,mm relative to each other along the $x$-$y$-$z$ diagonal. We also scanned the 20\,cm diameter spherical surface with 1,000 and 3,000 points and homogeneous coverage. Results were consistent in every case, so we settled for the double Cartesian grid.

We followed three different procedures for the shimming optimization: minimization via Mathematica, differential evolution with BBO, and creating a continuous-variable precursor with Ipopt and seeding it to BBO. In each, we studied the possibility of having positive-negative (i.e. +1,-1,0) or only-positive (+1,0) values, which we call $\alpha_j$ for the $j^\text{th}$ cube. For all procedures we pre-calculate an encoding matrix $M_{i,j}$ which gives the contribution of the $j^\text{th}$ cube at each point $i$ in the FoV. This means the total field is $B_\text{t}=B_0+M\alpha$, with $B_0$ the field before shimming. In Mathematica and BBO the cost function is the homogeneity in the FoV points, whereas for Ipopt we use the squared deviation $\sum_i [(B_\text{t})_i -\langle B_\text{t}\rangle]^2$, where $\langle B_\text{t}\rangle$ is the mean total field in the FoV.

\subsection{Gradients}
\label{sec:MetGrad}
We want to achieve image spatial resolutions of 1\,mm, gradient ramp times of \SI{500}{\micro s} and reach $k_\text{max}$ in $<2$\,ms, all with an inexpensive power amplifier delivering up to 15\,A and 15\,V. These numbers translate into a target gradient strength of 12\,mT/m, i.e. an efficiency of 0.8\,mT/m/A. To ease thermal management in case of operation with a more powerful amplifier, our gradient coils are based on water-jetted copper plates rather than wires (see Sec.\,\ref{sec:ResGrad}). As explained in Sec.\,\ref{sec:theory}, we follow the usual target method to optimize for gradient linearity \cite{turner86}, adding a cost term that limits power dissipation \cite{liGrad2008}. An inductance term is not included, but we later check consistency with the constraint of \SI{500}{\micro s} rise times.

Once the desired combination of basis functions is obtained in Matlab (The Mathworks Inc. Massachusetts, USA), we convert contours to wires, add a second wire at a distance of 1\,mm and keep the negative of the resulting region as the intended track. When turning contours to wires, we add an extra turn to the stream function contours, so that the negative has the specified number of turns. These tracks are then imported to COMSOL and we perform a magnetic simulation to obtain the inductance, and an electric simulation to obtain resistance. We then perform a transient simulation in LT Spice (Analog Devices, Inc., Massachusetts, USA) to determine minimum voltage required to ramp the gradients in $<\SI{500}{\micro s}$.

\subsection{Fabrication and assembly}
\label{sec:MetFab}
The 19 polyethylene rings of the main magnet, and the sockets for the cubic magnetic elements, have been machined by computer numerical control (CNC). Every ring is covered with a 2.5\,mm thick laser-cut aluminum plate machined to match the ring shape and fixed with screws. All rings are traversed by 16 inner and 16 outer threaded rods, for tight axial fastening.

Similarly, the shimming unit is composed of 23 polyethilene rings, with CNC machined sockets and threaded rods for fastening. Once inserted, the unit is fixed with screws to the main magnet.

The gradient module consists of four concentric elliptical parts. From the inside out we have the support structures for GX, GZ, GY and a protecting cover used for mechanical fixation. These parts are all 3D printed in polylactic acid (PLA). Every gradient coil lobe is a water-jetted copper plate curved with the help of an elliptical mould and later fixed to its corresponding 3D printed support with cyanoacrylate glue (glue drying is accelerated by a cyanoacrylate activator). We install a thermal sensor per lobe to monitor local temperatures and feed an interlock circuit if these are too high. The four concentric PLA parts are screwed together to integrate the gradient module.

An important concern is the repulsion force between magnet rings at assembly, which can limit bare-handed operability and require the use of special procedure for assembly, such as e.g. rails. We simulate the force between two rings in COMSOL, and check with a calculation of the dipole-dipole force between the collection of individual cubes in Julia.

\section{Results}
\subsection{Main magnet}

\subsubsection{Symmetries and other considerations.}
\label{sec:ResSymm}
The first configurations we explored aimed at displacing the imaging region towards one of the magnet mouths to ease patient access without sacrificing magnet length and, ideally, homogeneity. In principle, brain imaging would be possible with asymmetric ring configurations even with a 51\,cm long magnet, if the center of the FoV is shifted axially by 5\,cm. We tried to optimize a 51\,cm long magnet ($d_\text{m}=20$\,cm, $d_\text{M}=28$\,cm, $\Delta x=0.44$\,cm, 2 layers, 22 rings) following the methods described in Sec.\,\ref{sec:MetSymm} and \ref{sec:MetOpt}, shifting axially the center of the FoV by 5\,cm, both with the DEAP and BBO libraries. However, we did not find a configuration with homogeneity better than 9,000\,ppm in the 20\,cm diameter FoV. Since the asymmetry also makes it more sensitive to experimental tolerances and imperfections, the below results are all with axially symmetric magnets.

\begin{table}
\small
\begin{center}
\begin{tabular}{|l|l|l|l|l|l|}
\hline
Magnet configuration                  & \begin{tabular}[c]{@{}l@{}}$l=12$\,mm\\ 27 rings\\ 3 layers\end{tabular} & \begin{tabular}[c]{@{}l@{}}$l=12$\,mm\\ 27 rings\\ 4 layers\end{tabular} & \begin{tabular}[c]{@{}l@{}}$l=15$\,mm\\ 23 rings\\ 3 layers\end{tabular} & \begin{tabular}[c]{@{}l@{}}$l=19$\,mm\\ 19 rings\\ 2 layers\end{tabular} & \begin{tabular}[c]{@{}l@{}}$l=19$\,mm\\ 19 rings\\ 3 layers\end{tabular} \\ \hline
Field strength (mT)           & 74                                                                   & 100                                                                  & 99                                                                   & 92                                                                  & 141                                                                  \\ \hline
Homogeneity (ppm)      & 4,300                                                                 & 7,631                                                                 & 5,000                                                                 & 4,830                                                                & 10,600                                                                \\ \hline
Weight (kg) & 92                                                                   & 122                                                                  & 122                                                                  & 106                                                                 & 158                                                                  \\ \hline
\end{tabular}
\caption{Main magnet performance with different cube sizes and numbers of layers, where the bore has minor/major axes with $d_\text{m}=20$\,cm, $d_\text{M}=28$\,cm, and the total magnet length is $\approx 44$\,cm. The last row considers only the weight of the magnetic elements. }
\label{tab:MagConfigs}
\end{center}
\end{table}

%
%
We then tried to develop an intuition for the influence of the cube size, the number of layers in the rings and the number of rings in the assembly on the overall performance of $\approx44$\,cm long, axially symmetric configurations with $d_\text{m}=20$\,cm, $d_\text{M}=28$\,cm bore dimensions (Table\,\ref{tab:MagConfigs}). We consider the magnet acceptable if the resulting strength is $>80$\,mT,  homogeneous to $<5,000$\,ppm with magnetic elements amounting to around 100\,kg. The latter constraint rules out three of the optimized configurations. From the remaining two, the one with smaller elements (12\,cm) yields the best results in terms of homogeneity but the field is weaker than specified. As a rule of thumb, for a given magnet aspect ratio and given FoV extent, the optimal configuration seems to make use of the largest possible magnetic elements, as long as there are enough degrees of freedom to homogenize the field. We therefore ultimately decided for 19\,mm cubes arranged in two layers.

\begin{figure}
	\includegraphics[width=16cm]{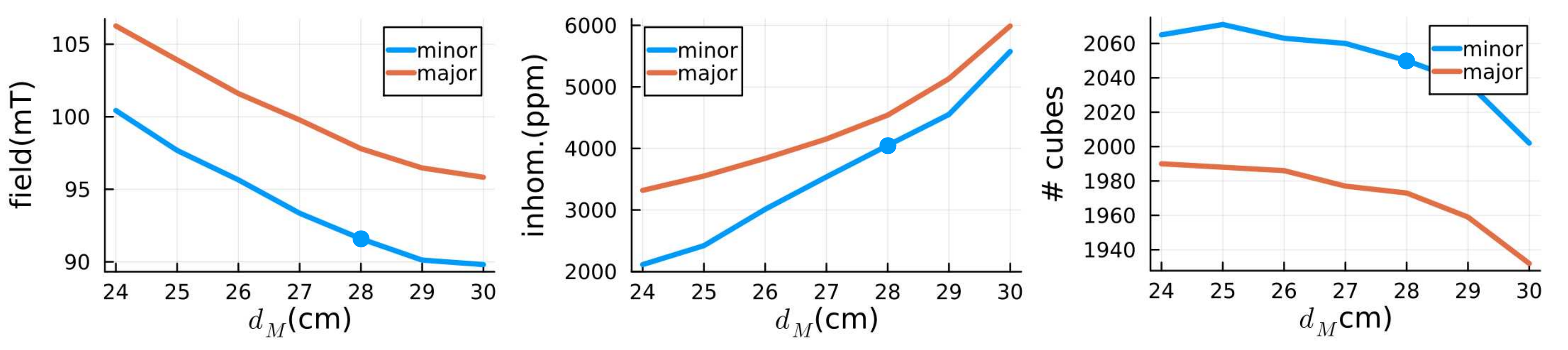}
	\caption{Field amplitude $B_y$ (left), inhomogeneity (middle) and number of magnetic cubes (right) as a function of major axis' diameter. Each point is the result of 200 averages of independent magnet optimizations for homogeneity at the given magnet parameters. Magnetic cubes are N48 grade and their size is 19mm. There are 19 rings with 2 layers each. We have highlighted with blue points the chosen configuration.}
	\label{fig:rMOpt}
\end{figure}

To select the size of the major bore axis $d_\text{M}$, we set $d_\text{m}=20$\,cm, $L=44.5$\,cm, $l=19$\,mm, $n_\text{L}=2$, $n_\text{R}=19$, $n_\text{M}=3$ and sought optimal configurations for $d_\text{M}$ values ranging from 24 to 30\,cm (Fig.\,\ref{fig:rMOpt}). Every optimization was run for 50,000 iterations of differential evolution with BBO with the default algorithm (Sec.\,\ref{sec:MetOpt}), and the number of cubes for each ring is allowed to vary between $n_\text{G}$ and $n_\text{M}$ compatible with a maximum magnet radius of $35$\,cm along the minor axis, excluding the $n_\text{M}$ mouth rings which are fixed at $n_\text{G}$. The full optimization used in Fig.\,\ref{fig:rMOpt} takes 170 seconds in an AMD Ryzen 9 7950X processor, single-threaded. The blue (red) curves correspond to the 2\,cm RF gap capacitors along the minor (major) axis. As expected, the field weakens and becomes less homogeneous for increasing $d_\text{M}$. The total number of magnetic cubes, and therefore the weight, does not vary as dramatically. When the gap capacitors lie along the major axis the field is stronger but less homogeneous, and the number of element decreases. Seeking a compromise between bore opening and homogeneity, we decided for $d_\text{M}=28$\,cm with capacitors along the minor axis (blue markers in Fig.\,\ref{fig:rMOpt}). This configuration yields $B_0\approx92$\,mT and 4,000\,ppm with $n_\text{C}=2,050$ elements.

\begin{figure}
	\includegraphics[width=16cm]{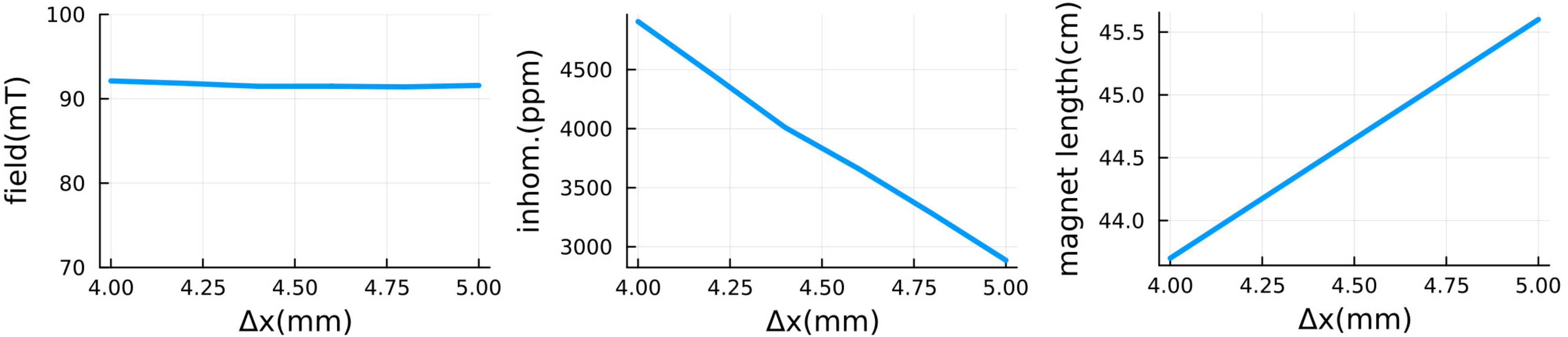}
	\caption{Field amplitude $B_y$ (left), inhomogeneity (middle) and total magnet length (right) as a function of inter-ring spacing. Each point is the result of 200 averages of independent magnet optimizations for homogeneity at the given magnet parameters. Magnetic cubes are N48 grade and their size is 19mm. There are 19 rings with 2 layers each.}
	\label{fig:deltaX}
\end{figure}

Figure \ref{fig:deltaX} shows the dependence of field strength, homogeneity and length on the spacing between consecutive rings, $\Delta x$ (also 50,000 iterations of differential evolution with BBO with the default algorithm; 200 averages, 70 seconds computation time in the same processor). The field strength is insensitive to $\Delta x$ in this range, but the homogeneity is strongly dependent. To stay at the 4,000 ppm level, we opted for $\Delta x=4.4$\,mm, which was convenient from a machining point of view due to the local availability of aluminum plates of this thickness for the ring covers, and furthermore yields a total magnet length $L\approx44.5$\,cm, barely above the 44\,cm deemed optimal in Sec.\,\ref{sec:MetDim}.

\subsubsection{Magnet dimensions.}
\label{sec:ResDim}
We analyze here the performance of different methods and libraries to determine the optimal arrangement and dimensions for an elliptical Halbach array with $l=19$\,mm, $d_\text{m}=20$\,cm, $d_\text{M}=28$\,cm, $n_\text{R}=19$ rings, $n_\text{L}=2$ layers per ring, with inter-ring spacing $\Delta x=4.4$\,mm, and total length of $L\approx 44.5$\,cm. 

Genetic algorithms as given by EvoLP (population of 25,000 individuals, 25 iterations) (Sec.\,\ref{sec:MetOpt}) ran for over 13 minutes (same processor), but were unable to find a configuration with homogeneity $\lesssim 4,300$\,ppm, and none of the executions seemed to converge. Among the algorithms available in the BBO library (see \cite{Feldt2018} for details), we found that an initial population of 50 (the default) yields the best results. Its differential evolution algorithms provided homogeneities down to 3,800\,ppm and often converged to the configuration we ended up selecting for fabrication. The only exception was the \texttt{de-rand2} variant, which was outperformed by the other DEAs. The natural evolution methods (\texttt{separable-nes}, \texttt{xnes}, \texttt{dxnes}) never reached $<4,000$\,ppm. The direct search algorithm \texttt{generating-set-search} worked reasonably, while \texttt{probabilistic-descent} never achieved $<6,000$\,ppm. Resampling memetic search was found to be extremely slow and never reached homogeneities below 4,200\,ppm. Stochastic Approximation never reached below 4,400\,ppm, and random-search was very fast but yielded inhomogeneities above 4,700\,ppm for all the population sizes used.

An interesting feature of the BBO library is that it provides multi-objective optimization, allowing to incorporate both homogeneity and field strength terms in the cost function. While here we dealt with field strength by careful design of number of layers and cube sizes, such parameters could be optimization variables in a more generic procedure where both field strength and homogeneity are maximized.

\begin{figure}
	\includegraphics[width=16cm]{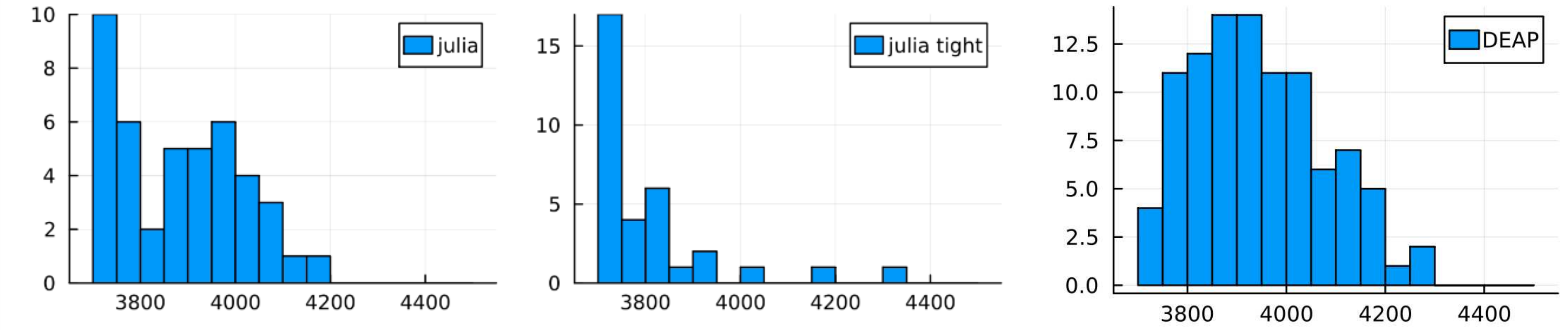}
	\caption{Frequency values of field homogeneity for 100 optimization runs, with bins in the range [3,700, 4,500]\,ppm in 50\,ppm steps, for the three optimization methods described in the main text: (left) BBO without fixing magnet number for outer rings, `julia'; (middle) BBO with fixed outer rings, `julia tight'; (right) DEAP without fixing outer rings.}
	\label{fig:conv}
\end{figure}

An important performance metric for any optimization is how frequently an algorithm reaches a good enough result. Figure\,\ref{fig:conv} shows histograms for 100 optimization runs of DEAP (25,000 population size, 250 iterations) and BBO (default method with 30,000 iterations), with the FoV as defined in the python DEAP script\cite{GenAlgRepo} (i.e. a dense Cartesian octant). For BBO we consider two possibilities for the three outer mouths rings: they are either optimization variables, within the range [$n_\text{G}$, $n_\text{MAX}$] rather than [$n_\text{S}$, $n_\text{MAX}$] (we call this option `julia'); or they are fixed to have $n_\text{G}$ magnets in the inner layer (`julia tight'). From the plot we observe that DEAP has similar accuracy than as `julia'. However, when we fix the outer rings (`julia tight') we significantly increase accuracy, highlighting the importance of physical intuition to constrain the space of configurations. We did not study this `tight' configuration with DEAP, so we can only conclude that both DEAP and differential evolution algorithms in BBO have similar accuracies for our problem. 

Regarding globality, we checked that the solutions with lowest inhomogeneity for the three cases in Fig.\,\ref{fig:conv} correspond to the same (apparently optimal) configuration [39,39,39,59,59,59,47,55,58,57] with 91\,mT and 3,800\,ppm, which we implemented in our prototype (Fig.\,\ref{fig:FinalMag}). Julia is a highly efficient programming language and this enabled us to check exhaustively the configuration space for the `tight' arrangement (optimization space of size $16^7$), finding that the resulting configuration is indeed the global optimum. 

The amount of FoV points included in the optimization strongly affects the computational efficiency. For the cases above we used a hollow sphere with 10\,cm radius and angles $\phi\in[0,\pi]$, $\theta\in[0,2\pi]$ were scanned with 8 steps (64 FoV points). We checked that increasing a factor 2 in each angle and running over possible radii $r\in[0,10]$\,cm in 9 steps indeed leads to the same magnet rings configuration. Following this procedure we also checked that the option `julia', also converged to the same configuration with many more FoV points.

The candidate configuration was simulated in COMSOL Multiphysics to estimate the relevance of magnetic coercitivity effects (see Sec.\,\ref{theo:opt}). The result was a field of 89\,mT and 3,900\,ppm, which is to be compared with 91\,mT and 3,800\,ppm in the dipole approximation. These numbers are sufficiently close to justify the value of our approach, especially since the massive number of configurations we have explored would have been out of reach with COMSOL. Finally, let us note that the COMSOL simulations show the 5-Gauss line is comparable to other low-field Halbach scanners and compatible with system portability and installation virtually anywhere \cite{OReilly2020,GuallartNaval2022}.

\subsubsection{Magnet.}
\label{sec:ResMag}

\begin{figure}
	\includegraphics[width=15cm]{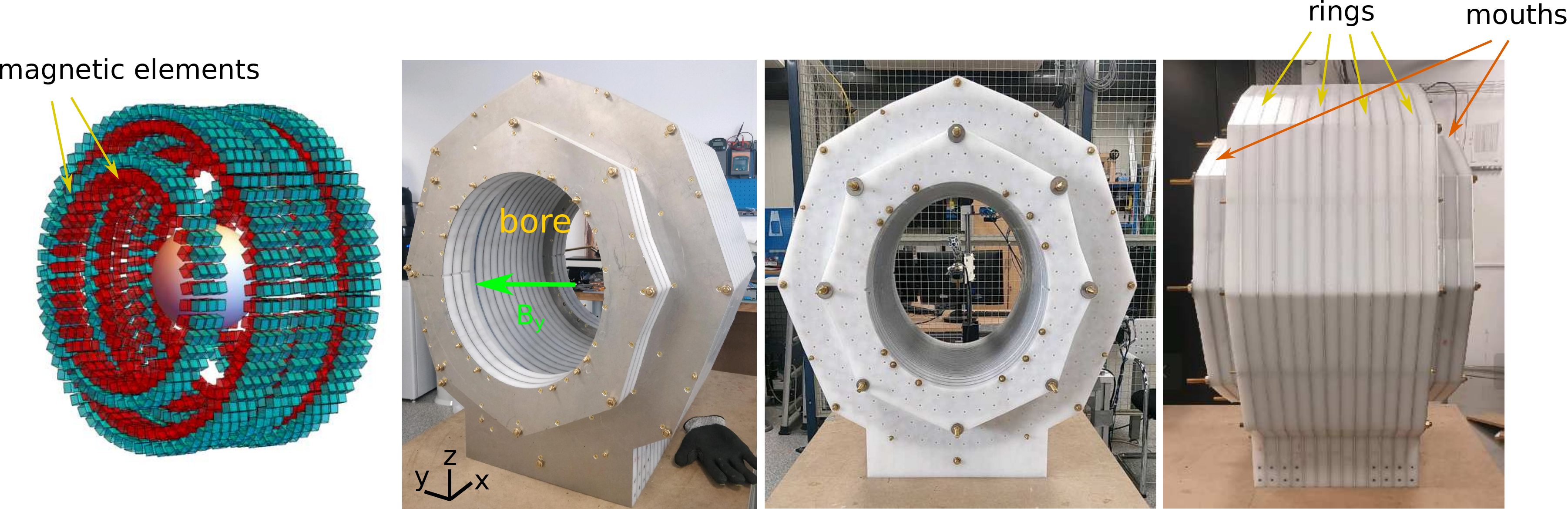}
	\caption{Assembled magnet. (left) Optimal configuration of magnetic cubic elements (red: inner layer, blue: outer layer) with the spherical FoV inside. (right) Different views of the assembled magnet with aluminum lids.}
	\label{fig:FinalMag}
\end{figure}

Following the above results, we built the final magnet (Fig.\,\ref{fig:FinalMag}) with rings made out of polyethylene, N48 neodymium cubic magnetic elements, and 2.5\,mm thick aluminum covers. The inter-ring spacing is $\Delta x=4.4$\,mm, the total magnet length is $L=44.5$\,cm, and the elliptical bore dimensions are $d_\text{m}=20$\,cm and $d_\text{M}=28$\,cm (Table\,\ref{tab:FinalMag}). 

\begin{table}[]
\begin{center}
\begin{tabular}{|l|l|l|l|l|l|l|l|}
\hline
$d_\text{m}$   & $d_\text{M}$   & l    & $n_\text{R}$ & $n_\text{M}$ & $n_\text{L}$ & $\Delta x$  & L  \\ \hline
20\,cm & 28\,cm & 19\,mm & 19 & 3  & 2  & 4.4\,mm & 44.5\,cm\\ \hline
\end{tabular}
\caption{Assembled configuration parameters.}
\label{tab:FinalMag}
\end{center}
\end{table}

The magnetic field was mapped with a THM1176 probe (Sec.\,\ref{sec:MetShim}). The field strength at the magnet center is 85.4\,mT, with 7,600\,ppm. This deviates significantly from the COMSOL simulations (89\,mT and 3,900\,ppm), making coercitivity an unlikely candidate to explain the disparity. On the other hand, to our best knowledge, all low-field Halbach magnets reported in the literature suffer a drop in field strength and homogeneity with respect to simulations \cite{OReilly2019,GuallartNaval2022,Wenzel2021}. Candidate explanations include positioning errors of the magnetic elements due to tolerances in the machining of the ring sockets combined with magnetic interactions between the cubes, tolerances in the assembly of the ring stack, or systematic deviations in the magnetization magnitude or orientation with respect to those specified by the vendors of magnetic elements (we have simulated random variations of multiple parameters, and none of those seem to lead to deviations as large as we experimentally observe). Indeed, the magnetization of some cubic elements we have measured separately seems to be weaker than specified by the manufacturer by around 5-10\,\%, a plausible cause for both the 5\,mT decrease in field at the center of the FoV and the loss of homogeneity. Regarding imperfections in the assembly of the ring stack, the simulated repulsion force between rings is sizable: 2.8\,kN for this magnet, compared to 340\,N in \cite{GuallartNaval2022}. This can lead to small but cumulative deformations in the rings, inducing also field deviations.

\subsection{Shimming unit}
\label{sec:ResShim}
Proper shimming of the main magnetic field with an extra set of Halbach rings requires detailed knowledge of the field in the first place (Sec.\,\ref{sec:MetShim}). We found that Cartesian maps with mesh resolution worse than 1\,cm resulted in different shimming configurations every time the main field was measured. After acquiring a second Cartesian map with 1\,cm mesh size, displaced 5\,mm diagonally along the $x-y-z$ direction, shimming configurations started to converge to the same final distribution. We opted for this double Cartesian scan, rather than a single Cartesian with isotropic mess size of 5\,mm, to cut the measurement time by a factor of 4 (8\,hours rather than 32). 

Mathematica's \texttt{NMinimize} function got stuck when the FoV included many points, both with continuous and integer variables, unless the mesh size was increased to 2\,cm, in which case computation times were $\approx30$\,min. Optimization from scratch with BBO also did not give satisfactory results, probably because the search space is too large (there are 2,001 sockets in the shimming unit, see Sec.\,\ref{sec:MetShim}). On the other hand, minimization of the square deviation function in Ipopt with continuous variables bounded by $|\alpha_i|<=1$ provided good precursor configurations. In a desktop computer, this takes around 8\,min, where over 7\,min are for RAM allocation of the problem and the optimization takes less than 30\,s. Memory usage by the JuMP framework (Sec.\,\ref{sec:MetOpt}) is massive (113\,GB) even reducing the FoV points to 1,154 in a hollow sphere of radius $9.5\leq r \leq 10$\,cm, although it can be done in similar times with Zram (a RAM compression library) and an extra swap space in a fast NVMe-type solid state disk, with only 65\,GB RAM. These precursors were given as seeds for BBO with field inhomogeneity as cost function (again with integer rounding of the variables), and were optimized during $2\cdot 10^6$ iterations in around 4\,min. This combination is convenient because the BBO optimization easily copes with the full FoV sphere and many more points, once Ipopt has provided the initial heavy lifting.

\begin{figure}[h!]
	\includegraphics[width=16cm]{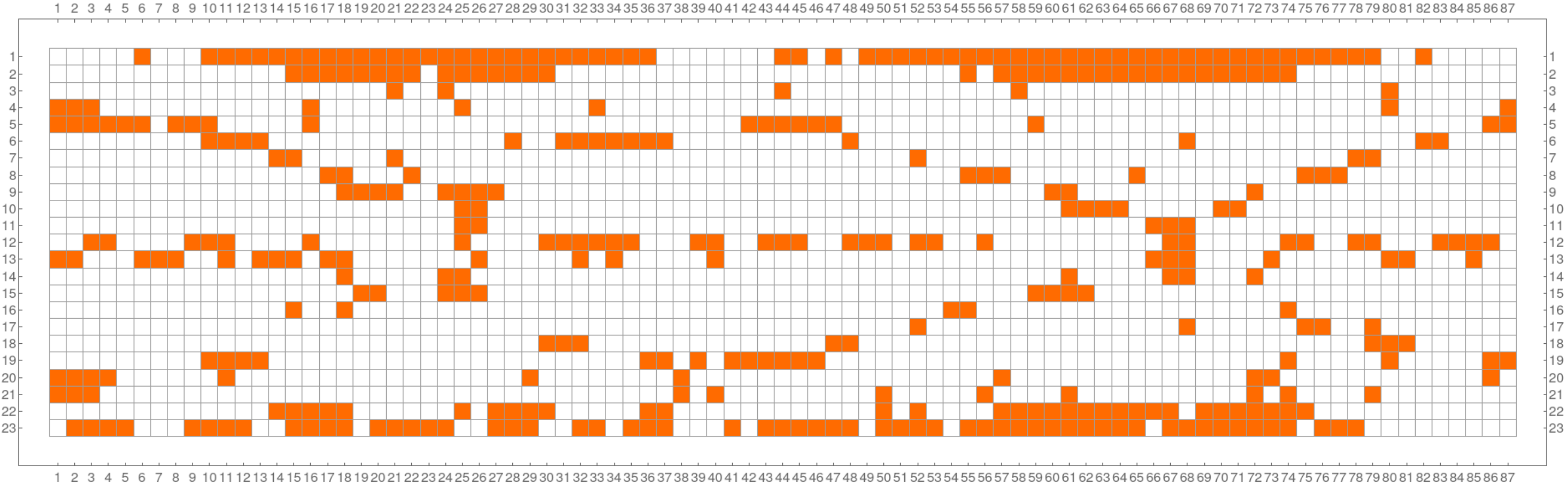}
	\caption{Only-positive shimming configuration obtained with Ipopt and BBO. Orange sockets have magnet cubes oriented along the Halbach direction, while white slots are empty. Vertical axis is ring number (increasing along $x$), and horizontal is slot position along the ring (with the first cube at ($d_\text{m}/2,0$), and increasing clock-wise).}
	\label{fig:FinalShim}
\end{figure}

An important decision is whether to take only-positive or negative-positive solutions (Sec.\,\ref{sec:MetShim}). Our conclusion is that the former yields worse homogeneity ($\sim$1,000\,ppm higher) but higher field strength ($\sim$4\,mT more) and less extra weight. All in all, we opted for the higher field with the optimal shimming in Fig.\,\ref{fig:FinalShim}. 

With the final shimming, the predicted field is 87\,mT and 3,600\,ppm, and we measured it to be 87\,mT with 5,100\,ppm. For completion, we also assembled a positive-negative solution which resulted in 83\,mT with 6,900\,ppm, compared to the predicted values of 82\,mT with 2,200\,ppm. Thus, only-positive configuration comes significantly closer to theoretical predictions, perhaps due to coercitivity effects becoming more prominent for pairs of magnetic elements with opposite signs. We also measured the shimming field on its own, outside of the magnet, noticing for both configurations that the field was 10\,\% weaker than expected from simulations, probably due to a slightly smaller remanence than specified by the manufacturer.

\subsection{Gradients}
\label{sec:ResGrad}

\begin{figure}[h!]
	\begin{center}\includegraphics[width=13cm]{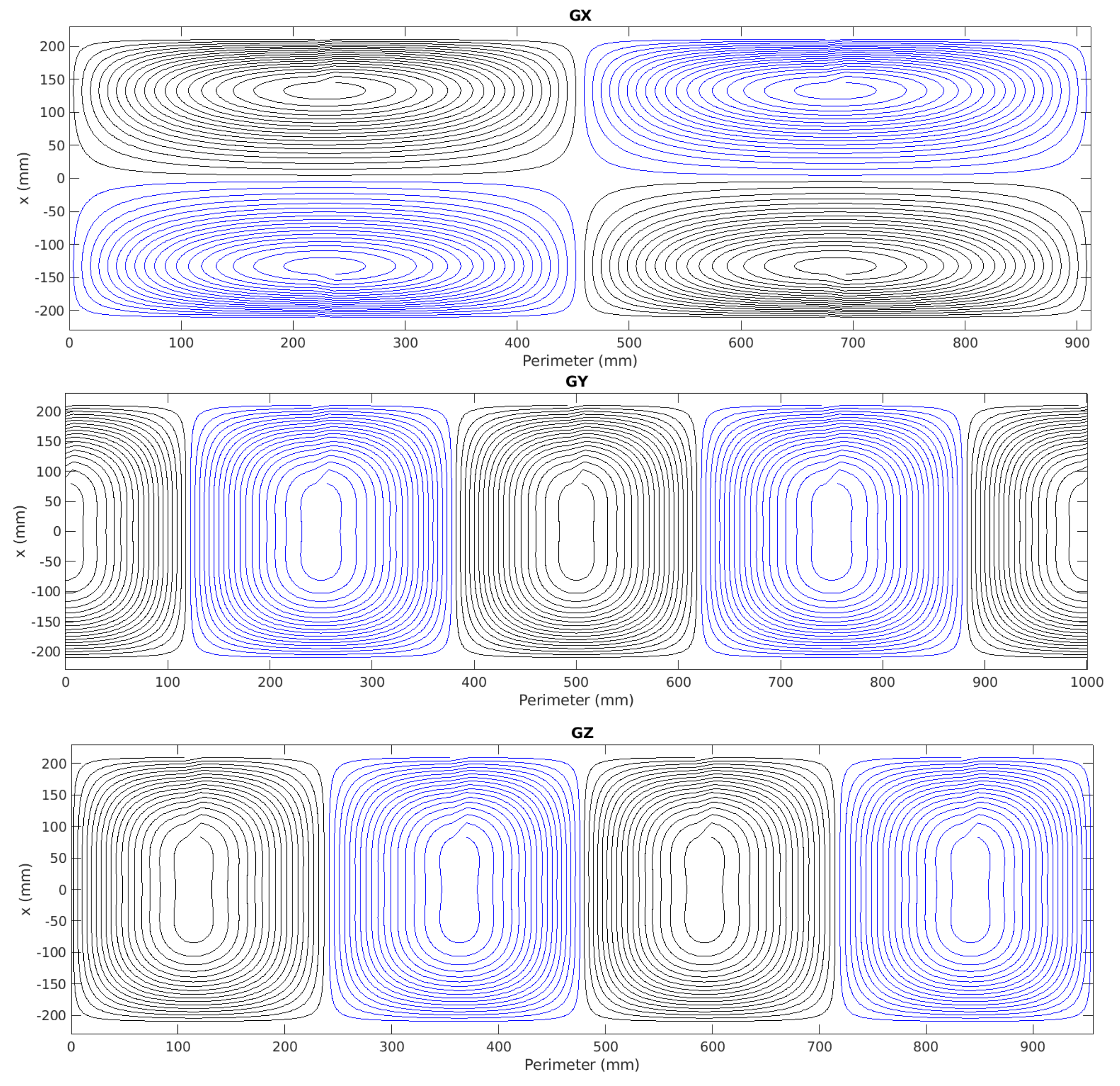}\end{center}
	\caption{Contours of optimal stream functions for each gradient, after joining contiguous contours. The negative of these contours is finally transformed into copper tracks. Note that in the GY gradient, black lobes (those along coordinate $y$) are shorter in perimeter due to the ellipticity of the support.}
	\label{fig:Stream}
\end{figure}

\begin{figure}[h!]
	\begin{center}\includegraphics[width=15cm]{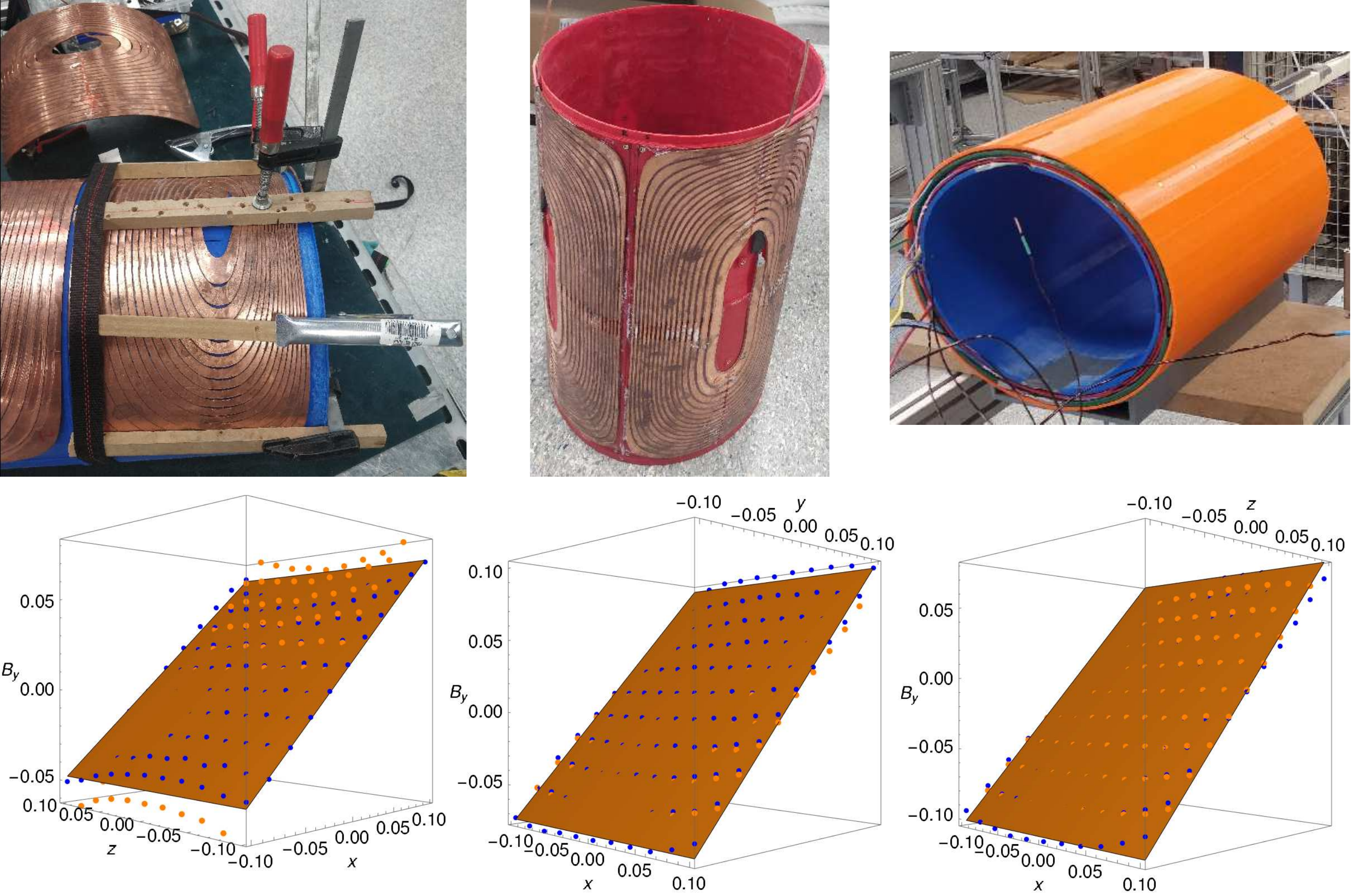}\end{center}
	\caption{Top.- Track-based elliptical gradients. (Left) GX gradient assembly process where metallic track is being fixed by cyanoacrylate glue to the support. (Middle) GZ gradient already assembled. (Right) Full set of assembled gradients with protection cover (orange). Bottom.- Measured gradient fields (points) vs ideal fields (orange plane) in a 20\,cm long cylinder of diameter 20\,cm. (Left) GX field measured at $y=0$ (blue) and at $y=6$\,cm (orange). (Middle) GY field measured at $z=0$ (blue) and at $z=6$\,cm (orange). (Right) GZ field measured at $y=0$ (blue) and at $y=6\,$cm (orange).}
	\label{fig:Grads}
\end{figure}

Following the target field method delineated in Sec.\,\ref{sec:MetGrad}, we obtained simulated gradient designs with efficiencies (0.61,0.87,0.87)\,mT/m/A for the GX, GY, GZ gradients, comparable to our previous scanner \cite{GuallartNaval2022}. The resistances are (88, 128, 112)\,mOhm, which are 3-4 times lower due to the use of a massive copper plate (2\,mm thick with 1\,mm wide water-jet cuts, see Fig.\,\ref{fig:Grads}) rather than wires. While inductances are a factor $\approx 1.5$ higher, $15$\,V suffice for ramp times $<\SI{500}{\micro s}$.
In Figure\,\ref{fig:Stream} we plot stream functions contours for each gradient in real space, where the GY gradient notably has different perimeter lenghts for different lobes, due to ellipticity.

Once assembled, we measure efficiencies and resistances very similar to those simulated. Deviations from linearity in a 20\,cm diameter sphere are also similar to the simulated (18, 13, 13)\,\%. In Figure\,\ref{fig:Grads} we plot gradient fields for different planes of interest. It must be noted that in our previous setup, with gradients restricted within the central magnet region, the GX gradient field had non-linearity of 20\,\% in a sphere of 15\,cm diameter, while in this setup it reduces to 9\,\% in 15\,cm, and 18\,\% in 20\,cm, highlighting the advantage of full-length gradient supports.

Aside from the elliptical cross-section, two relevant advantages of this design are the reduced power dissipation and, most importantly, the fact that the full gradient support results in an extension of the available axial FoV by 4.4\,cm with respect to the magnet in Ref.\,\cite{GuallartNaval2022}, even if the new magnet is significantly shorter (44.5\,cm vs 51\,cm).

\section{Discussion}

Designing a Halbach magnet involves deciding the general symmetries, the amount of relevant degrees of freedom, the tolerances, and the constraints on size, weight, field strength and homogeneity, among others. While one could envision a much more comprehensive and general method of design, such as a sarcophagus-like shape, industrial simplicity and computability of optimal configurations can impose harder constraints than other considerations. For example, any deviation away from longitudinal homogeneity, such as having different ellipticities for each ring, imposes a similar shape to gradient supports, forcing the designer to drop basis-function target field methods; and the same goes for RF design. Othere possibilities include not stacking up rings, having non-coaxial positioning of magnets (such as in patent \cite{patenteGirados}), or having magnetic units with more faces. However, they all lead to new challenges in manufacturing and assembly, and they introduce new sources of possible errors which can lead to field strength and homogeneity reductions by e.g. failure to implement the desired magnetic units orientation.

At any rate, even restricting to a configuration based on a stack of rings, improvements to our method are possible:
\begin{enumerate}
	\item We could have modified the Halbach rotation rule for better homogeneity (as in Tewari and Webb\cite{permanentWebb2023} with a single parameter $\tilde{m}$), or even leave magnetization angles as variables, which presumably leads to an extremely large configurational space.
	\item Assuming a set of possible magnetic cube sizes for different layers or even magnetic grades, though the former option could probably conflict with longitudinal tight-packing, because ring width would be misused by smaller cubes in a given layer.
	\item Using different numbers of layers for each ring, such as having more layers in mouths than in central rings. While we briefly explored this possibility, finding worse performance, we did not explore having central rings with smaller cubes, which could have enabled a better homogeneization. That is, fix very strong magnetic contributions from mouths by using large cubes and more layers, and use smaller cubes for fine tuning in the central rings. This option could be automatized by letting the algorithm choose number of rings and sizes of cubes (unique for each ring) while keeping the total length of the magnet fixed. We envision that this option would be amenable to optimization with the presented algorithms in Julia. A possible pitfall is an increase in weight, but this can be included in the cost function, either by weighted sum of terms, or by multi-objective optimization (as available in BBO library).
	\item An asymmetric bell-shaped longitudinal configuration could have been explored, with rings closing in above the head, and opening up near the shoulders, maybe even with more eccentric rings to accomodate for the shoulders. This would imply optimization of ring radii (as per $N_i$) and eccentricity, quadrupling the optimization variables (twice because it is asymmetric, twice because of eccentricities), which is not so demanding. However, simplicity of gradient supports would limit the extent of this asymmetry, since near the shoulders gradient lobes would tend to have currents more spatially compressed.
\end{enumerate}

A crucial aspect demonstrated here is that code optimization, library/language performance and accuracy of the optimization method can make the difference between dismissing a valid type of design and keeping it. In our case, the code is so fast that one can exhaustively check the full configuration space of the `julia tight' design. If the number of variables was to be quadrupled, this might mean the difference between reaching a near-optimal configuration or simply dismissing the design because the output of optimization does not attain the desired homogeneity of strength. While we have shown here that convergence of BBO and DEAP libraries is quite good, it can plausibly be the case that many near-optimal configurations are present, some of which could be advantageous in terms of assembly or stability against errors, and could be left out simply by ignorance of their existence. For example, two equally good solutions in theory could be very different in terms of their resilience against dispersion of magnetization strength and orientations. If the code is fast, we could in principle incorporate an automatized check against random variation of these parameters, dismissing the less-resilient configuration, maybe even as a cost function `term'.

It is also notable that the constraints on magnet size and weight, together with aiming at good homogeneity, have produced a configuration with inter-ring magnetic repulsion at the limit of human assembly without special tools, which is convenient.

\section{Conclusions and Outlook}
The toolbox described in this paper has allowed us to build a portable Halbach magnet with an elliptic bore amenable to heads and large extremities with full-length gradient support, with a magnetic field of 87\,mT @ 5,100\,ppm after shimming, and which fits through doors and weighs 216\,kg. This paves the way for future small length-to-ratio ergonomic and portable MRI magnets, and these methods could be extrapolated to full-body Halbach magnet design.

\section*{Contributions}

Main idea by FG, JA and JMB. Magnet opimization by FG. Mechanical design by EP with contributions from PGC, PM, JB, RB and FJ. Gradient optimization by TGN with contributions from JB, JMA. Magnet and gradient assembly, and measurements by EP, PGC, TGN, PM, JB, RB, FJ. The paper was written by FG and JA with input from all authors.

\section*{Conflict of interests}
JMA, JB, JMB, FG and JA have a patent pending on elliptical Halbach magnets. PM and FJ are employed at PhysioMRI Tech S.L., and JB is employed at Tesoro Imaging S.L. JMA, JMB, FG and JA are co-founders of PhysioMRI Tech S.L. All other authors declare no competing interests.


\section*{Acknowledgment}

This work was supported by the European Innovation Council under grant 101136407, Ministerio de Ciencia e Innovación of Spain under grant PID2022-142719OB-C22, Generalitat Valenciana under grant CIPROM/2021/003 and Agencia Valenciana de la Innovación under grants INNVA1/2022/4 and INNVA1/2023/30.

\vspace{1 cm}
\bibliography{myrefs}

\providecommand{\newblock}{}
\begin{thebibliography}{10}
\expandafter\ifx\csname url\endcsname\relax
  \def\url#1{{\tt #1}}\fi
\expandafter\ifx\csname urlprefix\endcsname\relax\def\urlprefix{URL }\fi
\providecommand{\eprint}[2][]{\url{#2}}

\bibitem{Deoni2022}
Deoni S~C, Medeiros P, Deoni A~T, Burton P, Beauchemin J, D'Sa V, Boskamp E, By
  S, McNulty C, Mileski W, Welch B~E and Huentelman M 2022 {\em Scientific
  Reports 2022 12:1\/} {\bf 12} 1--9 ISSN 2045-2322
  \urlprefix\url{https://www.nature.com/articles/s41598-022-09760-2}

\bibitem{GuallartNaval2022}
Guallart-Naval T, Algar{\'{i}}n J, Pellicer-Guridi R, Galve F, Vives-Gilabert
  Y, Bosch R, Pall{\'{a}}s E, Gonz{\'{a}}lez J, Rigla J, Mart{\'{i}}nez P,
  Lloris F, Borreguero J, Marcos-Perucho {\'{A}}, Negnevitsky V,
  Mart{\'{i}}-Bonmat{\'{i}} L, R{\'{i}}os A, Benlloch J~M and Alonso J 2022
  {\em Scientific Reports 2022 12:1\/} {\bf 12} 1--11 ISSN 2045-2322
  \urlprefix\url{https://www.nature.com/articles/s41598-022-17472-w}

\bibitem{Algarin2023}
Algar\'in J, Guallart-Naval T, Gastaldi-Orqu\'in E, Bosch R, Lloris F, Pall\'as
  E, Rigla J, Martínez P, Borreguero J, Alamar R, Mart\'i-Bonmat\'i L,
  Benlloch J, Galve F and Alonso J 2023 {\em arXiv preprint\/}
  \urlprefix\url{http://arxiv.org/abs/2303.09264}

\bibitem{Sheth2021}
Sheth K~N, Mazurek M~H, Yuen M~M, Cahn B~A, Shah J~T, Ward A, Kim J~A, Gilmore
  E~J, Falcone G~J, Petersen N, Gobeske K~T, Kaddouh F, Hwang D~Y, Schindler J,
  Sansing L, Matouk C, Rothberg J, Sze G, Siner J, Rosen M~S, Spudich S and
  Kimberly W~T 2021 {\em JAMA Neurology\/} {\bf 78} 41--47 ISSN 2168-6149

\bibitem{Sheth2022}
Sheth K~N, Yuen M~M, Mazurek M~H, Cahn B~A, Prabhat A~M, Salehi S, Shah J~T, By
  S, Welch E~B, Sofka M, Sacolick L~I, Kim J~A, Payabvash S, Falcone G~J,
  Gilmore E~J, Hwang D~Y, Matouk C, Gordon-Kundu B, Rn A~W, Petersen N,
  Schindler J, Gobeske K~T, Sansing L~H, Sze G, Rosen M~S, Kimberly W~T and
  Kundu P 2022 {\em Scientific Reports 2022 12:1\/} {\bf 12} 1--11 ISSN
  2045-2322 \urlprefix\url{https://www.nature.com/articles/s41598-021-03892-7}

\bibitem{Liu2021}
Liu Y, Leong A~T~L, Zhao Y, Xiao L, Mak H~K~F, Chun A, Tsang O, Lau G~K~K,
  Leung G~K~K, Wu E~X and Linfang X 2021 {\em Nature Communications 2021
  12:1\/} {\bf 12} 1--14 ISSN 2041-1723

\bibitem{Sarracanie2020}
Sarracanie M and Salameh N 2020 {\em Frontiers in Physics\/} {\bf 8} 172 ISSN
  2296-424X

\bibitem{Wald2020}
Wald L~L, McDaniel P~C, Witzel T, Stockmann J~P and Cooley C~Z 2020 {\em
  Journal of Magnetic Resonance Imaging\/} {\bf 52} 686--696 ISSN 1522-2586

\bibitem{Webb2023}
Webb A and Obungoloch J 2023 {\em Nature 2023 615:7952\/} {\bf 615} 391--393
  ISSN 0028-0836

\bibitem{chonqing2020}
He Y, He W, Tan L, Chen F, Meng F, Feng H and Xu Z 2020 {\em Journal of
  Magnetic Resonance\/} {\bf 319} 106829 ISSN 1090-7807
  \urlprefix\url{https://www.sciencedirect.com/science/article/pii/S1090780720301476}

\bibitem{Soltner2010}
Soltner H and Bl{\"{u}}mler P 2010 {\em Concepts in Magnetic Resonance Part A:
  Bridging Education and Research\/} {\bf 36} 211--222 ISSN 15466086

\bibitem{OReilly2020}
O'Reilly T, Teeuwisse W~M, Gans D, Koolstra K and Webb A~G 2020 {\em Magnetic
  Resonance in Medicine\/}  mrm.28396 ISSN 0740-3194

\bibitem{gluckstern1983}
Gluckstern R~L and Holsinger R~F 1983 {\em IEEE Transactions on Nuclear
  Science\/} {\bf 30} 3623--3626

\bibitem{lund1996}
Lund S~M and Halbach K 1996 {\em Fusion Engineering and Design\/} {\bf 32-33}
  401--415 ISSN 0920-3796 proceedings of the Seventh International Symposium on
  Heavy Ion Inertial Fusion
  \urlprefix\url{https://www.sciencedirect.com/science/article/pii/S0920379696004966}

\bibitem{kustler2010}
Kustler G 2010 {\em IEEE Transactions on Magnetics\/} {\bf 46} 3601--3607

\bibitem{gluckstern1996}
Domigan P, Hass M~A and Gluckstern R Full brick construction of magnet assembly
  having a central bore

\bibitem{permanentWebb2023}
Tewari S and Webb A 2023 {\em Scientific Reports\/} {\bf 13} 2774

\bibitem{Cooley2018}
Cooley C~Z, Haskell M~W, Cauley S~F, Sappo C, Lapierre C~D, Ha C~G, Stockmann
  J~P and Wald L~L 2018 {\em IEEE Transactions on Magnetics\/} {\bf 54} 1--12

\bibitem{OReilly2019}
O'Reilly T, Teeuwisse W and Webb A 2019 {\em Journal of Magnetic Resonance\/}
  {\bf 307} 106578 ISSN 10907807
  \urlprefix\url{https://linkinghub.elsevier.com/retrieve/pii/S1090780719302174}

\bibitem{interior2000}
Potra F~A and Wright S~J 2000 {\em Journal of Computational and Applied
  Mathematics\/} {\bf 124} 281--302 ISSN 0377-0427 numerical Analysis 2000.
  Vol. IV: Optimization and Nonlinear Equations
  \urlprefix\url{https://www.sciencedirect.com/science/article/pii/S0377042700004337}

\bibitem{DiffEvol1997}
Storn R and Price K 1997 {\em Journal of Global Optimization\/} {\bf 11}
  341--359

\bibitem{DEcomparative2004}
Vesterstrom J and Thomsen R 2004 A comparative study of differential evolution,
  particle swarm optimization, and evolutionary algorithms on numerical
  benchmark problems {\em Proceedings of the 2004 Congress on Evolutionary
  Computation (IEEE Cat. No.04TH8753)\/} vol~2 pp 1980--1987 Vol.2

\bibitem{DEneural2020}
Baioletti M, Di~Bari G, Milani A and Poggioni V 2020 {\em Mathematics\/} {\bf
  8} ISSN 2227-7390 \urlprefix\url{https://www.mdpi.com/2227-7390/8/1/69}

\bibitem{DEneuralTopology2022}
Belciug S 2022 {\em Computers in Biology and Medicine\/} {\bf 146} 105623 ISSN
  0010-4825
  \urlprefix\url{https://www.sciencedirect.com/science/article/pii/S0010482522004152}

\bibitem{turner86}
Turner R 1986 {\em Journal of Physics D: Applied Physics\/} {\bf 19} L147
  \urlprefix\url{https://dx.doi.org/10.1088/0022-3727/19/8/001}

\bibitem{liGrad2008}
Li X, Xie D, Wang J and Zhang X 2008 Design of finite size uniplanar gradient
  coil for fully open mri system with horizontal magnetic field {\em 2008 World
  Automation Congress\/} pp 1--4

\bibitem{QIMS5938}
Chilla G~S, Tan C~H, Xu C and Poh C~L 2015 {\em Quantitative Imaging in
  Medicine and Surgery\/} {\bf 5} ISSN 2223-4306
  \urlprefix\url{https://qims.amegroups.org/article/view/5938}

\bibitem{BOLLMANN2021}
Bollmann S and Barth M 2021 {\em Progress in Neurobiology\/} {\bf 207} 101936
  ISSN 0301-0082 how high spatiotemporal resolution fMRI can advance
  neuroscience
  \urlprefix\url{https://www.sciencedirect.com/science/article/pii/S030100822030191X}

\bibitem{litzWireGrad2021}
Jia F, Littin S, Amrein P, Yu H, Magill A~W, Kuder T~A, Bickelhaupt S, Laun F,
  Ladd M~E and Zaitsev M 2021 {\em Journal of Magnetic Resonance\/} {\bf 331}
  107052 ISSN 1090-7807
  \urlprefix\url{https://www.sciencedirect.com/science/article/pii/S1090780721001415}

\bibitem{dipole2013}
Petruska A~J and Abbott J~J 2013 {\em IEEE Transactions on Magnetics\/} {\bf
  49} 811--819

\bibitem{streamFunc}
Pissanetzky S 1992 {\em {Measurement Science and Technology}\/}  667

\bibitem{humano}
J P and M Z 1979 {\em Human Dimension and Interior Space A Source Book of
  Design Reference Standards\/} (Whitney library of Design,Watson-Guptill
  Publications, New York)

\bibitem{webbIsmrm}
O'Reilly T and Webb A 2021 {\em {Proc. Intl. Soc. Mag. Reson. Med.}\/}  4041

\bibitem{GenAlgRepo}
{Thomas O'Reilly} 2018 Genetic algorithm optimization
  \url{https://github.com/LUMC-LowFieldMRI/HalbachOptimisation}

\bibitem{Julia}
Bezanson J, Edelman A, Karpinski S and Shah V~B 2017 {\em SIAM Review\/} {\bf
  59} 65--98

\bibitem{Feldt2018}
Feldt R 2018 Blackboxoptim.jl
  \url{https://github.com/robertfeldt/BlackBoxOptim.jl}

\bibitem{Sanchez-DiazEvoLP2023a}
Sánchez-Díaz X~F~C and Mengshoel O~J 2023
  \urlprefix\url{https://ceur-ws.org/Vol-3431/}

\bibitem{Wachter2006}
W\"achter A and Biegler L 2006 {On the Implementation of a Primal-Dual Interior
  Point Filter Line Search Algorithm for Large- Scale Nonlinear Programming}

\bibitem{JuMP}
Lubin M, Dowson O, {Dias Garcia} J, Huchette J, Legat B and Vielma J~P 2023
  {\em Mathematical Programming Computation\/} {\bf 15} 581–589

\bibitem{Wenzel2021}
Wenzel K, Alhamwey H, O'Reilly T, Riemann L~T, Silemek B and Winter L 2021 {\em
  Frontiers in Physics\/} {\bf 9} 704566 ISSN 2296424X

\bibitem{patenteGirados}
Vidarsson L Magnetic resonance imaging ({MRI}) system and method {US10018694B2}

\end{thebibliography}


\end{document}